\documentclass[manuscript]{aastex}
\usepackage{graphicx}
\usepackage{amssymb,amsmath}
\usepackage{color}
\definecolor{Blue}{rgb}{0.3,0.3,0.9}

\shorttitle{Uncertainties in Spectral Models}
\shortauthors{Bautista et al.}

\begin{document}
\title{Uncertainties in Atomic Data and Their Propagation 
Through Spectral Models. I.}

\author{M.A.~Bautista\altaffilmark{1}, V. Fivet\altaffilmark{1,2},
P. Quinet\altaffilmark{2}
 J.~Dunn\altaffilmark{3}, T.R.~ Gull\altaffilmark{4}, T.R. Kallman\altaffilmark{5},
C. Mendoza\altaffilmark{6} }
%},
%Ehab ElHoussieny\altaffilmark{1}, 
%        M.A.~Bautista\altaffilmark{1},
%        T.R.~Kallman\altaffilmark{3}}

\altaffiltext{1}{Department of Physics, Western Michigan University, Kalamazoo, 
MI 49008, USA \email{manuel.bautista@wmich.edu}
} 
\altaffiltext{2}{Astrophysique et Spectroscopie, Universit\'e de Mons – UMONS, 7000 Mons, Belgium and
IPNAS, Universit\'e de Li\`ege, Sart Tilman, 4000 Li\`ege, Belgium}
\altaffiltext{3}{Physical Science Department, Georgia Perimeter College, Dunwoody, GA 30338, USA} 
\altaffiltext{4}{Code 667, NASA Goddard Space Flight Center, Greenbelt, MD 20771, USA}
\altaffiltext{5}{Code 662, NASA Goddard Space Flight Center, Greenbelt, MD 20771, USA} 
\altaffiltext{6}{Centro de F\'{i}ısica, Instituto Venezolano de Investigaciones Cient\'{\i}ficas (IVIC), PO Box
20632, Caracas 1020A, Venezuela}

%==================================================================================
%
\begin{abstract}
We present a method for computing uncertainties in spectral models, i.e.
level populations, line emissivities, and emission line ratios, based upon the
propagation of uncertainties originating from atomic data.
We provide analytic expressions, in the form of linear
sets of algebraic equations,  for the coupled uncertainties
among all levels. These equations can be solved efficiently for any set of 
physical conditions and uncertainties in the atomic data.
We illustrate our method applied to spectral models of O III
and Fe II and discuss the impact of the uncertainties 
on atomic systems under different physical conditions. 
As to intrinsic uncertainties in theoretical atomic data, we
propose that these uncertainties can be estimated from the
dispersion in the results from various independent calculations.
This technique provides excellent results 
for the uncertainties in A-values of forbidden transitions 
in [Fe II]. 
\end{abstract}
%
%==================================================================================
%
\section{Introduction}
Much can be learned about the behavior and evolution 
of an astronomical source through analysis of properly calibrated spectra. Interpretation can lead to estimates of
density and temperature conditions, 
the chemical composition, the dynamics, and the
sources of energy that power the emitting object. Such an interpretation  
requires 
modeling, with sufficiently high accuracy, the excitation and ionization balance of plasmas out of local 
thermodynamic equilibrium (LTE).
But ultimately, the accuracy of the models
depends on the quality of the atomic/molecular data employed. 

At present, atomic data exists for most spectral lines observed from the
infrared to the X-rays. These data account for most processes leading to tens of thousands of transitions from all ionic stages of nearly all
elements of the first five rows of the periodic table. However, this huge amount of data has been obtained primarily through
theoretical calculations with only sparse checks against experimental measurements.

Despite many advances in spectral modeling, mostly in terms of increased completeness and improved quality of atomic/molecular parameters,
a generalized quantitative estimate of uncertainties of the resultant models has not been provided.  In recent years, a few authors have presented methods
based on the Monte Carlo numerical technique for propagating uncertainties 
through spectral models, e.g. \cite{wesson2012,ballance}. 
As these techniques  are very inefficient, 
 their general applicability to complex         
spectral modeling is very limited. Finding a general and efficient method for estimating uncertainties in spectral models is
important for two reasons.
First, the accuracy of atomic/molecular data must be known before reliable conclusions can be provided on physically realistic comparisons between theoretical and observed spectra. At present researchers
can only provide best fits to observed spectra without much understanding
of the uncertainties impacting the results. 
Second, homogeneously accurate atomic data for all transitions of a complex and/or very large atomic system, 
like for example systems with multiple
metastable levels (e.g. Fe II) or models with hundreds of energy levels as those needed in UV and X-ray spectroscopy, cannot be obtained.
In such models, error
propagation analysis of the spectrum could discriminate between a few critically important atomic transitions and the
very large numbers of less consequential transitions. Conversely, detailed error analysis could direct further theoretical and/or 
experimental efforts to selectivelly obtain specific atomic measures that would significantly improve spectral models, rather than trying to determine all possible rates at once.

This paper is organized as follows. Section 2 presents the analytical solution to the uncertainties
in level populations of a non-LTE spectral model for assumed uncertainties in
atomic parameters. In Section 3 we propose a mechanism to estimate
the uncertainties in atomic/molecular data and we test this by the
case of Fe II through
extensive comparisons with observed spectra. In Section 4 we discuss the uncertainties in line emissivities 
and emission line ratio diagnostics.
Section 5 presents our conclusions. 
For the sake of clarity, the rest of the paper deals explicitly with the case of population balance by electron impact excitation followed by
spontaneous radiative decay. It is also assumed that the plasma is optically thin. However,  we note that 
our method can easily be extended  
to ionization balance computations, to additional excitation mechanisms such as continuum and Bowen fluorescence, 
and to optically thick transitions.

\section{Uncertainties in Level Populations and Column Densities}

Under steady-state balance the population, $N_i$, of a level $i$ is given by 
\begin{equation}
N_i={\sum_{k\ne i}N_k(n_e q_{k,i} + A_{k.i})\over n_e \sum_{j\ne i} q_{i,j} + \sum_{j<i} A_{i,j}},
\end{equation}\label{popeq}
where $n_e$ is the electron density, $A_{k,i}$ is the Einstein spontaneous radiative rate from level $k$ to level $i$ and 
$q_{k,i}$ is the electron impact transition rate coefficient for transitions from level $k$ to level $i$. Here, we assume 
that the electron velocity distribution follows the Maxwell-Boltzmann function, thus $q_{k,i}$ and $q_{i,k}$ are both
proportional to a symmetrical effective collision strength, $\Upsilon_{k,i}=\Upsilon_{i,k}$, which is the source of
uncertainty in the collisional transition rates.
Assuming that the spectral 
model is arranged in increasing level energy order $A_{k,i}=0$ whenever $k<i$. We note that the second term
in the denominator of the above equation is the inverse of the lifetime of 
level $i$, i.e., $\tau_i=(\sum_{j<i} A_{i,j})^{-1}$.
This is important because lifetimes are generally dominated by a few strong transitions, which are much more accurately
determined than the weak transitions. Thus, $\tau_i$ carries smaller uncertainties than individual A-values. Then, equation (1)
can be written as
\begin{equation}
N_i={\sum_{k\ne i} N_k(n_e q_{k,i} + A_{k.i})\over 
n_e \tau_i \sum_{j\ne i} q_{i,j} + 1} \tau_i
=
{\sum_{k\ne i} N_k(n_e q_{k,i} + A_{k.i})\over n_e/n_i^c+ 1} \tau_i, 
\end{equation}
where $n_i^c$ is the so-called critical density of level $i$ and is defined as 
$n^c_i = (\tau_i \sum_{j\ne i} q_{i,j})^{-1}$.

The uncertainty in the population of level $i$, $\delta N_i$, can be computed as
\begin{equation}
\begin{split}     
(\delta N_i)^2 = \sum_{k\ne i}\left[\left({\partial N_i\over \partial \Upsilon_{k,i}}\right)^2 (\delta \Upsilon_{k,i})^2 +  
\left({\partial N_i\over \partial A_{k,i}}\right)^2 (\delta A_{k,i})^2 \right]
\\     +
\sum_{j\ne i} 
\left({\partial N_i\over \partial A_{i,j}}\right)^2 (\delta A_{i,j})^2 +
\sum_{k\ne i} \left({\partial N_i\over \partial N_k}\right)^2(\delta N_k)^2.
\end{split} 
\end{equation}\label{leverr}
The first three terms on the right hand side of this equation represent direct propagation of uncertainties from atomic rates
to or from level $i$. The last term in the equation correlates the uncertainty in level $i$ with the uncertainties in the level populations of all other
levels that contribute to it .
Then,
\begin{equation}
\begin{split}
\left({\delta N_i\over N_i}\right)^2
- \sum_{k\ne i} N_k^2{(n_eq_{k,i}+A_{k,i})^2\over \kappa^2} \left({\delta N_k\over N_k}\right)^2
=
\\ 
{1\over \kappa^2}\left[ n_e^2\sum_{k\ne i}(N_kq_{k,i}-N_iq_{i,k})^2 
\left({\delta \Upsilon_{k,i}\over \Upsilon_{k,i}}\right)^2 
+ \sum_{k> i}(N_kA_{k,i})^2\left({\delta A_{k,i}\over A_{k,i}}\right)^2
+ \left(
{N_i^2\over \tau^2}\right)^2 \left({\delta \tau_i\over \tau_i}\right)^2\right]
\end{split} 
\end{equation}\label{leverr2}
where 
$\kappa_i = \sum_{k\ne i} N_k (n_eq_{k,i}+A_{k,i}).$

This linear set of equations yields the uncertainties in the populations of all levels. Before proceeding to solve
these equations it is worth pointing out some important properties: (1) uncertainties are obtained relative to the
computed level populations regardless of the normalization adopted for these. This is important because while
some spectral models compute population relative to the ground level other models solve for normalized populations
such that $\sum N_k$ is either 1 or the total ionic abundance. Though, the equation above is generally applicable regardless of the normalization adopted.
(2) In the high density limit, $n_e\to \infty$, the right hand side of the equation goes to zero, thus the 
population uncertainties naturally go to zero as the populations approach the Maxwell-Boltzmann values (LTE conditions).
(3) By having an analytical expression for the propagation of uncertainties one can do a detailed analysis of the 
spectral model to identify the key pieces of atomic data that determine 
the quality of the model for any plasma conditions. 
(4) The set of linear equations for the uncertainties needs to be solved
only once for any set of conditions and the system is of the same size
as that for the level populations. This is unlike Monte Carlo approaches 
that require solving population balance equations hundreds of times,
which makes real-time computation of uncertanties 
impractical. 

The set of equations above can be readily solved by writing them as
\begin{equation}
B \bar x = \bar b,
\end{equation}
where $x_i = (\delta N_i/N_i)^2$, and the matrix and vector elements of $B$ and $\bar b$ are given by the equation \ref{leverr2}.

Figure 1 shows the populations and population uncertainties for the first four 
excited levels of O III as a function of the electron density at a temperature 
of $10^4$ K. For this computation we have assumed 5\% uncertainties in the lifetimes,
10\% uncertainties in individual A-values, and 20\% uncertainties in the 
effective collision strengths. The levels considered here are 
$2p^2\  ^3P_{0,1,2}$, $^1D_2$, and $^1S_0$. It is seen that levels 2 through 5 
have maximum uncertainties, 
$\sim$20\%, in the low density limit where the populations are determined by collisional excitations from the
ground level. As the electron density increases thermalization of levels with similar energies and radiative cascades
start becoming more important, which diminishes the contribution of uncertainties in collision strengths and enhances the importance of 
uncertainties in A-values. For high densities all population 
uncertainties naturally go to zero as the populations approach the Boltzmann limit.
Another thing to notice is that, the population uncertainties 
exhibit multiple contributions and peaks as the metastable levels $^3P_1$ and $^3P_2$ become populated
and the uncertainties in these propagate through higher levels. 

\begin{figure*}
\epsscale{0.7}\plotone{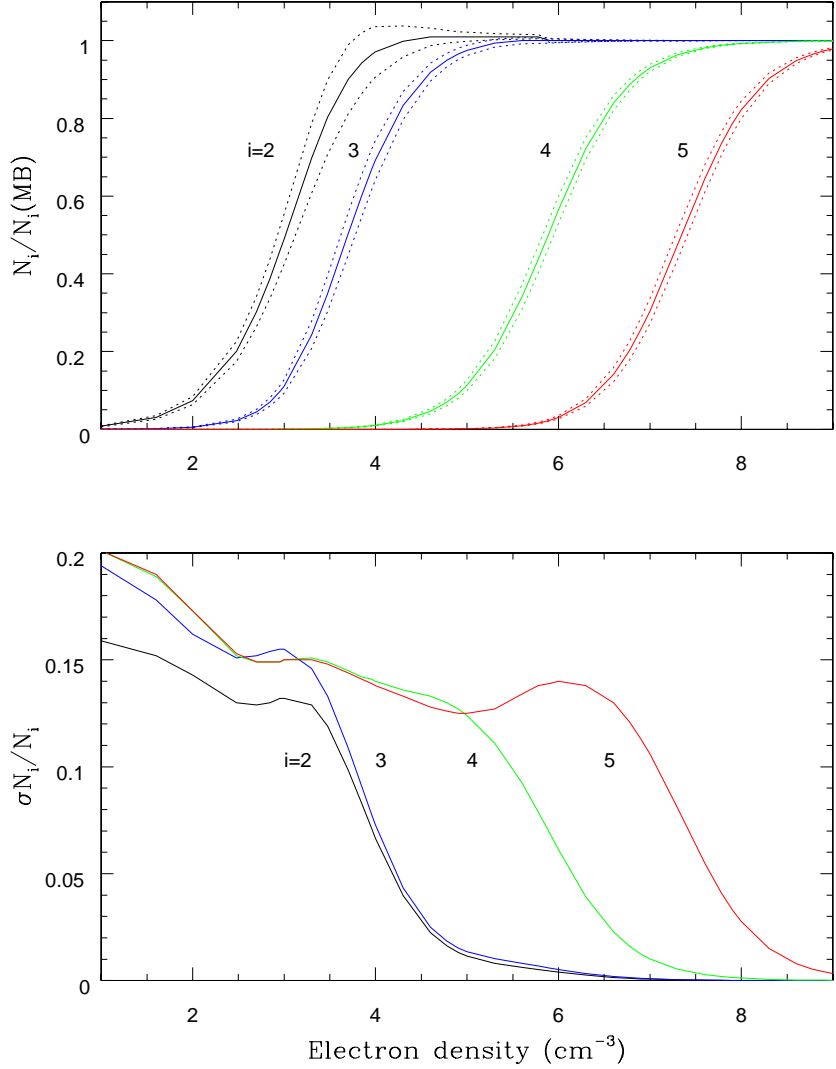}
\caption{Level populations relative to their Boltzmann limits (upper panel) and relative level population uncertainties (lower panel) 
for the $2p^2\ ^3P_1$ ($i=2$), $^3P_2$ ($i=3$), $^1D_2$ ($i=4$), and $^1S_0$ ($i=5$) excited levels of O III.}
\label{o3error}
\end{figure*}
Figure 2 shows the populations, relative to the ground level, and population 
uncertainties for the first eight excited levels of Fe II as a function of the 
electron density at a temperature of $10^4$ K. For these calculations we use 
atomic data as in \cite{bautista98} and assume uncertainties of 5\% in the lifetimes,
10\% in individual A-values, and 20\% in the effective collision strengths. The levels considered here are $3d^64s\ ^5D_{9/2,7/2,5/2,3/2}$ and $3d^7\ ^4F_{9/2,7/2,5/2,3/2}$. An interesting characteristic of the Fe II system
is that the $3d^7\ ^4F_{9/2}$ excited level is more populated, at least
according to the atomic data adopted here,  than the ground level at densities around $10^4$ cm$^{-3}$,
typical of H II regions. Moreover, under these conditions only $\sim$20\% of the total Fe II abundance is in the ground
level. This means that unlike lighter species, where excitation is dominated by the ground level or the ground multiplet,
in Fe II all metastable levels are strongly coupled and uncertainties in atomic data are expected to propagate in a 
highly non-linear fashion.
\begin{figure}
\epsscale{0.7}\plotone{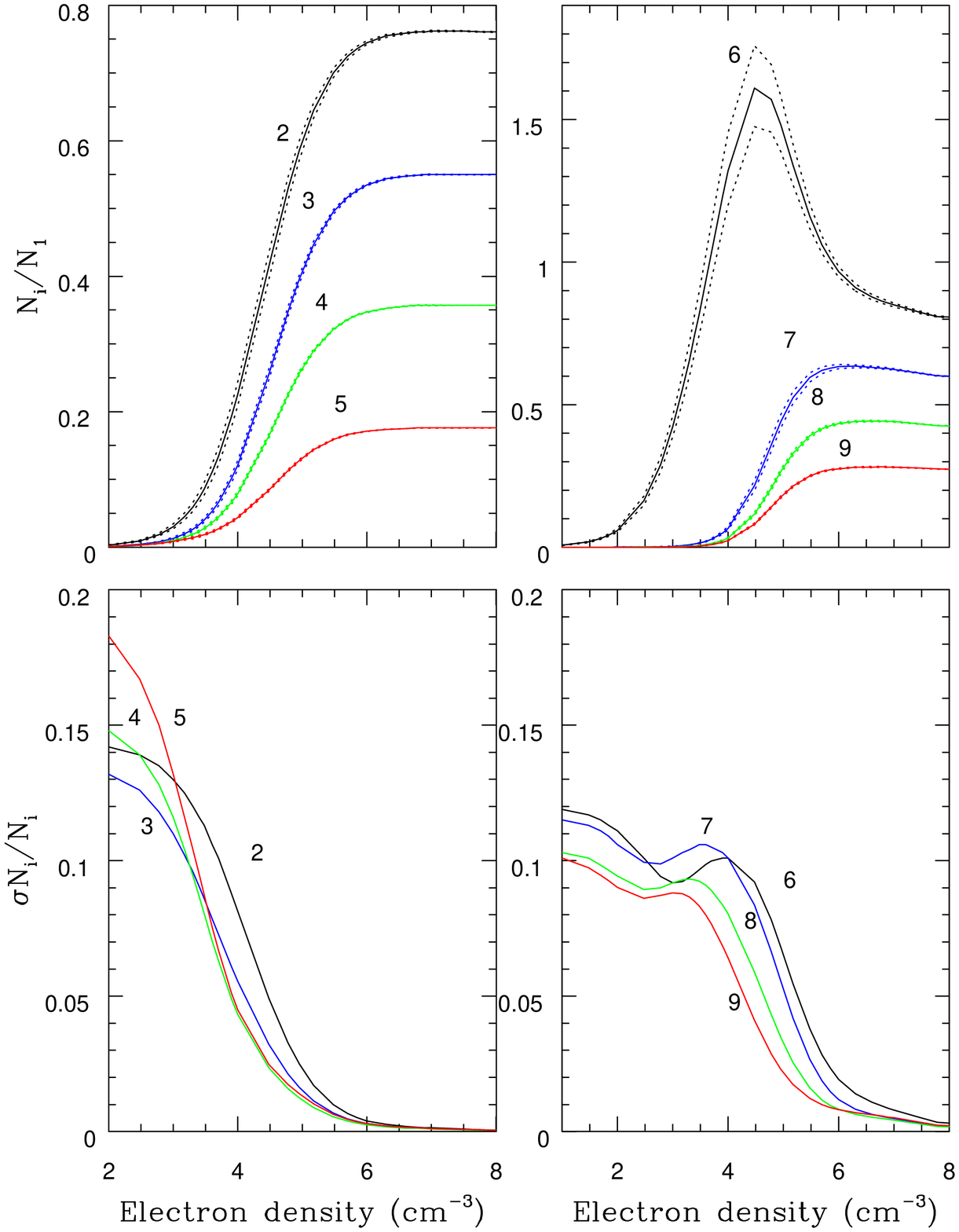}
\caption{Level populations relative to the ground level (upper panels) and relative level population uncertainties 
(lower panels) for the $3d^64s ^6D_{7/2}$ ($i=2$), $^6D_{5/2}$ ($i=3$), $^6D_{3/2}$ ($i=4$), $^6D_{1/2}$ ($i=5$),
$3d^7\ ^4F_{9/2}$ ($i=6$),  $3d^7\ ^4F_{7/2}$ ($i=7$), $3d^7\ ^4F_{5/2}$ ($i=8$), and $3d^7\ ^4F_{3/2}$ ($i=9$).}  
\label{errorfe2a}
\end{figure}

In Figure 3 we present the population errors for the lowest 52 levels of Fe II at $T_e=10^4$ K and $n_e=10^4$
cm$^{-3}$. These are all even parity metastable levels, except for the ground level. 
The figure shows the total estimated uncertainties together with the direct contributions from
uncertainties in  the collision strengths and A-values (first and second terms on the right hand side
of equation \ref{leverr}) and the contribution from level uncertainty coupling. It is observed  
that the collision strengths are the dominant source of uncertainty for all levels except level 6
($a\ ^4F_{9/2}$). For this level the uncertainty is dominated by the A-values and uncertainty couplings 
with levels of its own multiplet and levels of the ground multiplet. This is important because
we find that the $a\ ^4F_{9/2}$ level makes the largest contribution to
 the uncertainties in 36 of the lowest 52 levels of Fe II.
Unfortunately, the atomic data for the $^4F_{9/2}$ level are among the most uncertain parameters of the
whole Fe II system, as we discuss in the next section.

\begin{figure}
\epsscale{0.7}\plotone{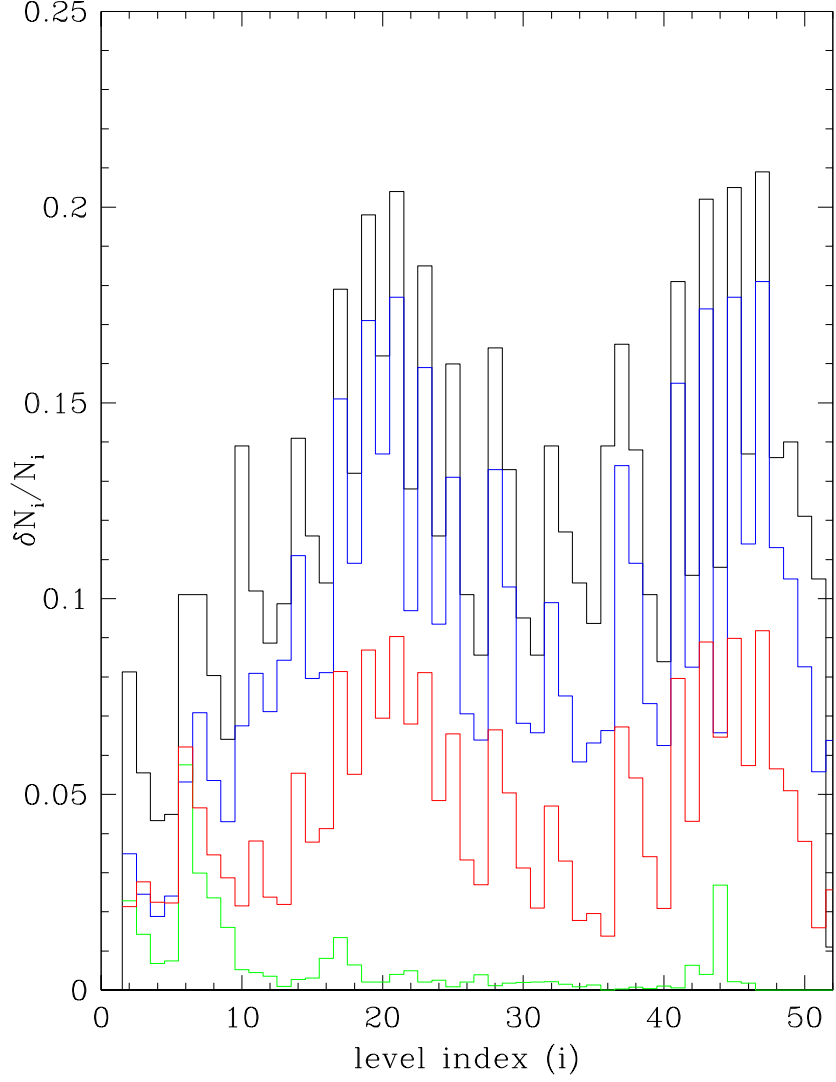}
\caption{Estimated level population uncertainties for the lowest 52 levels of Fe II at $T_e=10^4$ K and $n_e=10^4$ (black line).
Here we assume uncertainties for lifetimes, A-values, and collisional rates of 5\%, 10\%, and 20\% respectively.
The figure also depicts the contributions from uncertainties in collisional rates (blue line), A-values (green line), and
coupling of uncertainties among all levels (red line).} 
\label{errordist}
\end{figure}

\section{Estimating true uncertainties in atomic data}

In the previous section we adopted general uncertainties for lifetimes, A-values for forbidden transitions,
and effective collision strengths of 5\%, 10\%, and 20\%, respectively. In absence of generally accepted 
procedures to estimate uncertainties in theoretical atomic data, these kind of numbers are often 
cited in the literature as general guidelines; however, uncertainty estimates on specific rates are rarely
provided. In \cite{bautista09} we proposed that uncertainties in gf-values could be estimated from the
statistical dispersion among the results of multiple calculations with different methods and by different authors.
The uncertainties can be refined by comparing with experimental or spectroscopic data whenever available,
although these also have significant associated uncertainties. This approach is similar to what has been
done for many years by the Atomic Spectroscopy Data Center at the Nation Institute of Standards and Technology (NIST; http://www.nist.gov/pml/data/asd.cfm)
in providing a 'critical compilation' of atomic data.

In estimating uncertainties from the dispersion of multiple results one must keep in mind some caveats:
(a) Small scatter among rates is obtained when the computations converge to a certain value, yet such
a convergence is dependant on the maximum size of the quantum mechanical representation
treatable at the time of the computation. Thus, there is no
guarantee that every seemingly converged result is indeed correct, as some
values may
result from local minima in the parameter space.
(b) Large scatter among different calculations is expected in atomic rates where configuration interaction
and level mixing lead to cancellation effects. The
magnitude of these effects depends on the wave-function representation adopted. 
Thus, some computations maybe a lot more accurate than others for certain
transitions and if we knew which computation is the most 
accurate, then the scatter among all different computations may overestimate the true uncertainty.
However, detailed information about configuration and level mixing for
every transition is rarely available in the literature. 
Nevertheless, in absense of complete information about every transition rate
from every calculation, a 
critical comparison between the results of different calculations and other sources 
of data, if available, provides a reasonable estimate of the uncertainty in atomic/molecular rates.

Are the statistical dispersion values realistic uncertainty estimates? To answer this question we look at 
the intensity ratios between emission lines from the same upper level as obtained from observed astronomical spectra and
theoretical predictions. The advantage of looking at these ratios is that they depend only on the A-values,
regardless of the physical conditions of the plasma. Thus, the ratios ought to be the same in any spectra of
any source, provided that the spectra have been corrected for extinction. Fe II yields the richest spectrum 
of all astronomically abundant chemical species. Thus, high resolution optical and near-IR [Fe II]
lines are the best suited for the present experiment. 

One hundred thirty seven [Fe II] lines are found in the HST/STIS 
archived spectra of the Weigelt blobs of $\eta$ Carinae. 
Six medium dispersion spectra ($R$=6000 to 10,000) of the blobs were recorded between 1998 and 2004 at various orbital
phases of the star's 5.5-year cycle. Seventy eight [Fe II] lines are also present in the deep echelle spectrum ($R$=30~000)
of the Herbig-Haro object (HH~202) in the Orion nebula from \cite{mesadelgado}.
The importance of having multiple spectra from different sources and different instruments must not be overlooked.   
Multiple measurements of the same line ratio minimize the likelihood of systematic errors due to unidentified blends,
contamination from stellar emission,
and instrumental effects. 

\begin{figure}
\epsscale{0.7}\plotone{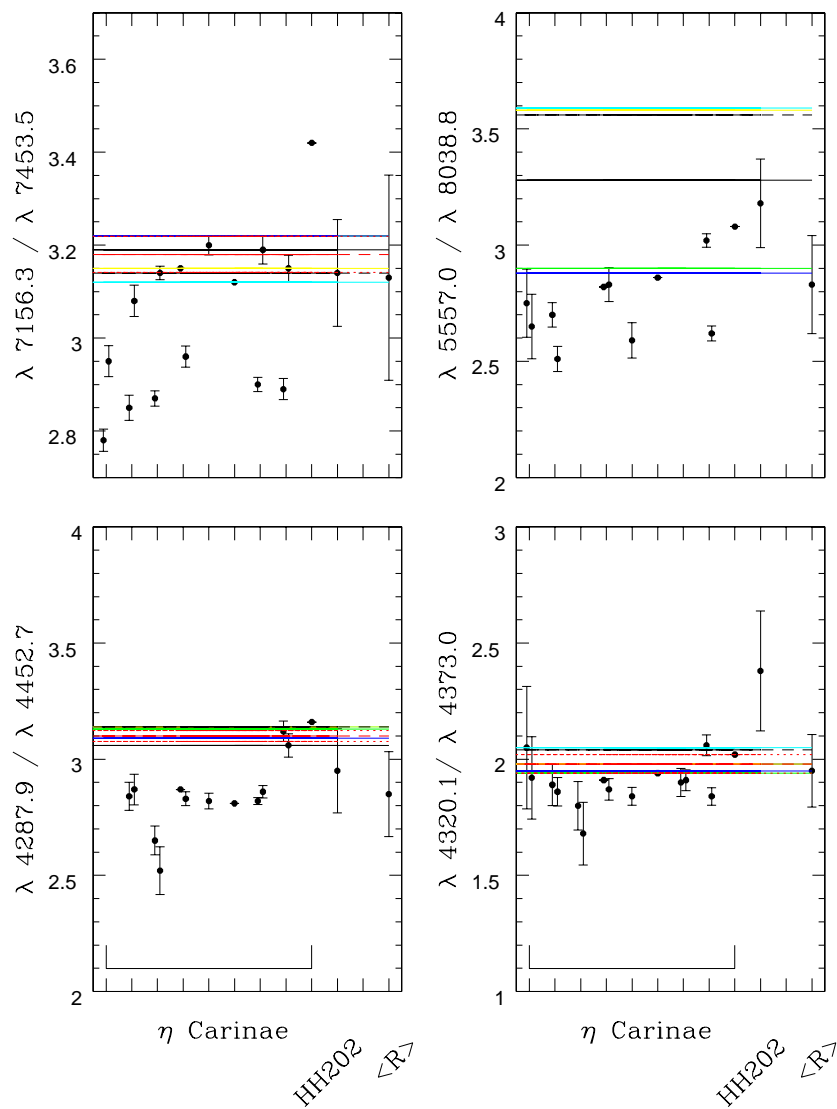}
\caption{Emission line ratios from transitions from the same upper level. The first nine points from left to right
results from our measured intensities in the HST/STIS spectra of the Weigelt blobs of $\eta$ Carinae. The tenth 
point is the measured ratio in the echelle spectrum of HH 202. The last point to the right depicts the average of all 
measurements and uncertainties given by the standard deviation. The horizontal lines represent the predictions from
several different computations of A-values.} 
\label{measure}
\end{figure}

From the observations, there are 107 line ratios reasonably well measured from the spectra. The
ratios are defined as
\begin{equation}
{\rm ratio} = max(F_1,F_2)/min(F_1,F_2),
\end{equation}
where $F_1$ and $F_2$ are the measured fluxes of two lines from the same upper level. Here, it is important
that the minimum of the two fluxes is put in the denominator for the ratio. Thus, the line ratios are 
unconstrained and they are all equally weighted when comparing with theoretical expectations.
Figure \ref{measure} illustrates a few line ratio determinations from several measurements
from spectra of $\eta$ Carinae and HH~202, as well as from various theoretical 
determinations. In practice, 
we perform up to four measurements of every observation for different spectral extractions
along the CCD and different assumptions about the continuum and the noise levels. 
Thus, we see that the scatter between
multiple measurements of a given ratio greatly exceed the statistical uncertainties in the line flux integrations.
Moreover, the scatter between measured line ratios often exceeds the scatter between theoretical predictions.
Full details about the Fe~II spectra and measurement procedures will be presented in a forthcoming paper, where
we will also present our recommended atomic data for Fe~II.

\begin{figure}
%\centerline{\includegraphics[angle=-90]{uncerfig.eps} 
\epsscale{0.7}\plotone{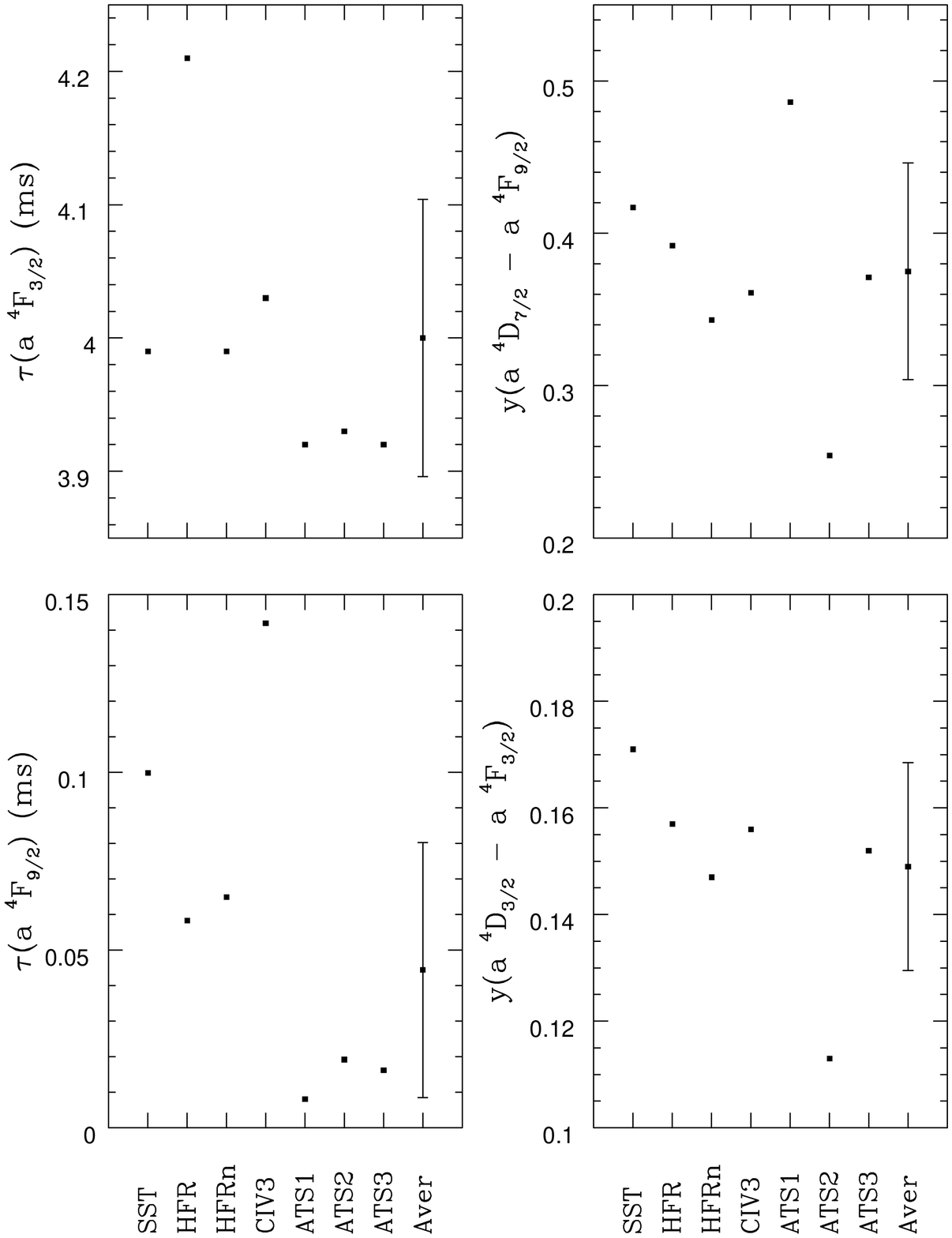} 
\caption{Theoretically calculated lifetimes and transition yields in Fe II. The calculations depicted are
SST: SUPERSTRUCTURE computation by \cite{quinet96}; HFR: HFR calculation also by Quinet et al.;   
HFRn: our new HFR calculation; CIV3: results by \cite{deb11}; ATS21, ATS2, and ATS3: our new AUTOSTRUCTURE 
calculations that extend over Quinet et al. The last point to the right of each panel depicts the average
value of the various determinations. The uncertainty bars for this point are set by the statistical dispersion 
between all values.} 
\label{uncerfig}
\end{figure}

For the present work we consider seven different computations of A-values for Fe II. These are the SUPERSTRUCTURE
and relativistic Hartree-Fock (HFR) calculations by \cite{quinet96}, the recent CIV3 calculation of
\cite{deb11}, and various new HFR and AUTOSTRUCTURE calculations that extend over previous works. Figure \ref{uncerfig}
presents a sample of theoretically calculated lifetimes and transition yields in Fe II. The yields are
defined as $y_{i,j}= A_{i,j}\times \tau_i$.
From the dispersion among various results, the average uncertainty in lifetimes
for all levels of the $3d^7$ and $3d^64s$ configuration is 13\%. More importantly, it is found that the uncertainty
in the critically important $a\ ^4F_{9/2}$ level is $\sim 80\%$, due to cancelation effects in the configuration
interaction representation of the  $a\ ^4F_{9/2} - a\ ^6D_{9/2}$ transition.

\begin{figure}
\epsscale{0.8}\plotone{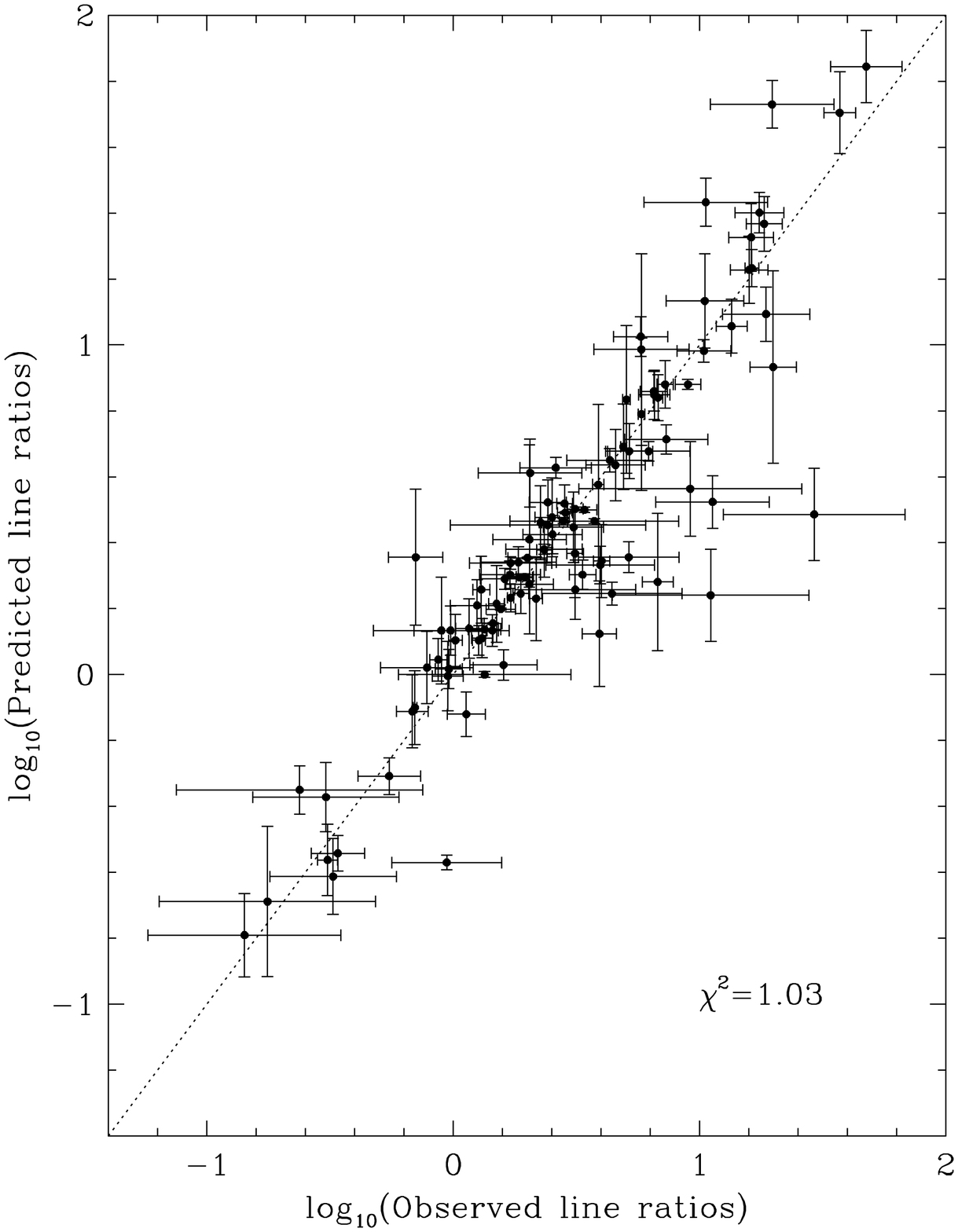}
\caption{Line ratios between line from the same upper level measured from optical nebular spectra vs. theoretical predictions.} 
\label{aratios}
\end{figure}

We compared the observed line ratios described above with the predictions from different sets of theoretical A-values.
Without uncertainty estimates for the theoretical values, the reduced-$\chi^2$ values from these comparison range from
2.2 to 3100 for the different sets of A-values. On other hand, if one adopts average A-values from all calculations
and uncertainties from the resultant standard deviations the reduced-$\chi^2$ is 1.03. This is indicative of
well estimated uncertainties, neither underestimated nor overestimated, and within these uncertainties there is
good agreement between theoretical and experimental line ratios. 
The comparison between observed and theoretical line ratios, including uncertainties, is presented in 
Figure \ref{aratios}.

\begin{figure}
\epsscale{0.7}\plotone{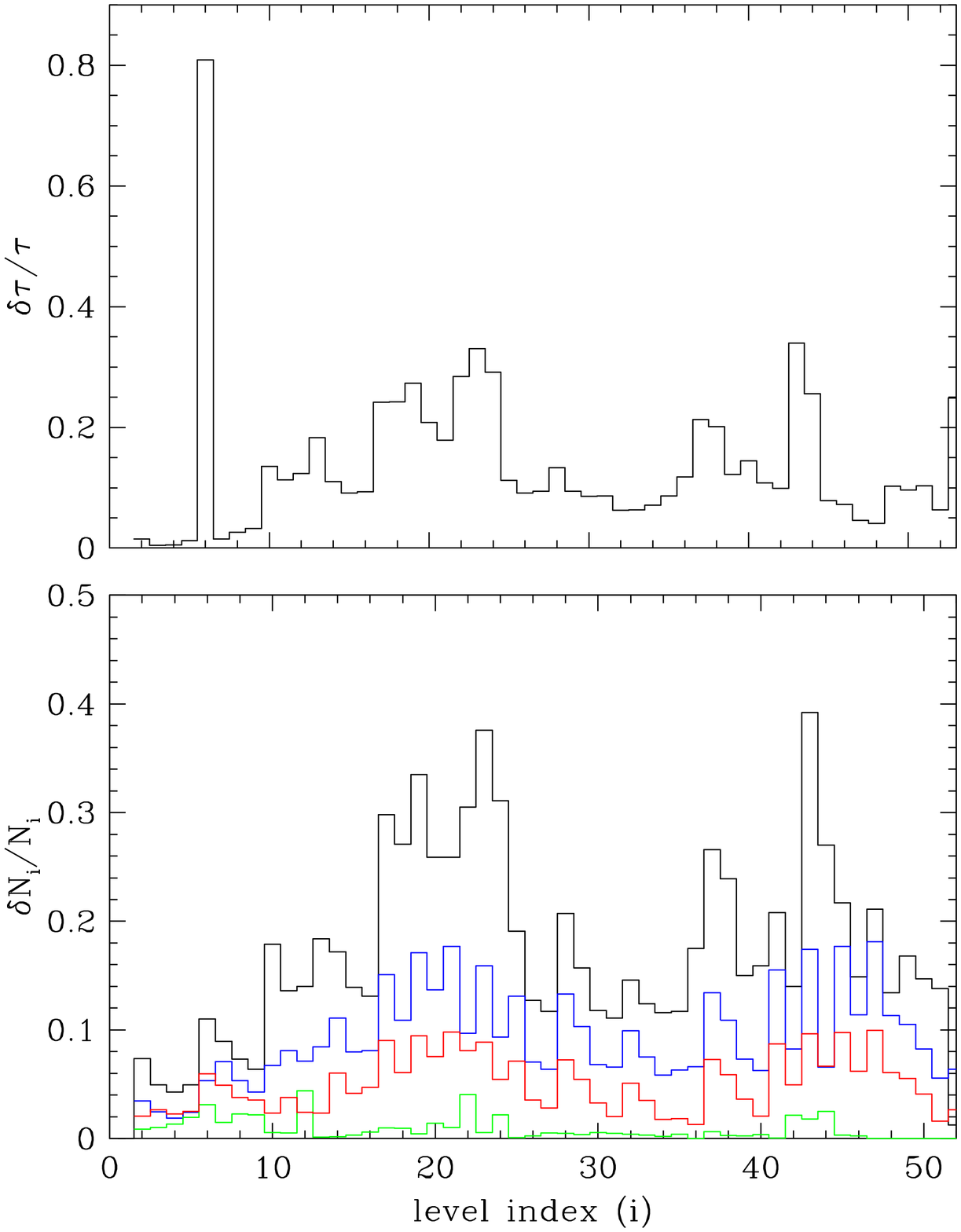}
\caption{The upper panel presents the estimated uncertainties in lifetimes
for the lowest 52 levels of Fe II. The lower panel is like Figure~\ref{errordist} but from uncertainties in lifetimes and radiative yields estimated from
the dispersion among various calculations.} 
\label{errordist2} 
\end{figure}

\begin{figure}
\epsscale{0.7}\plotone{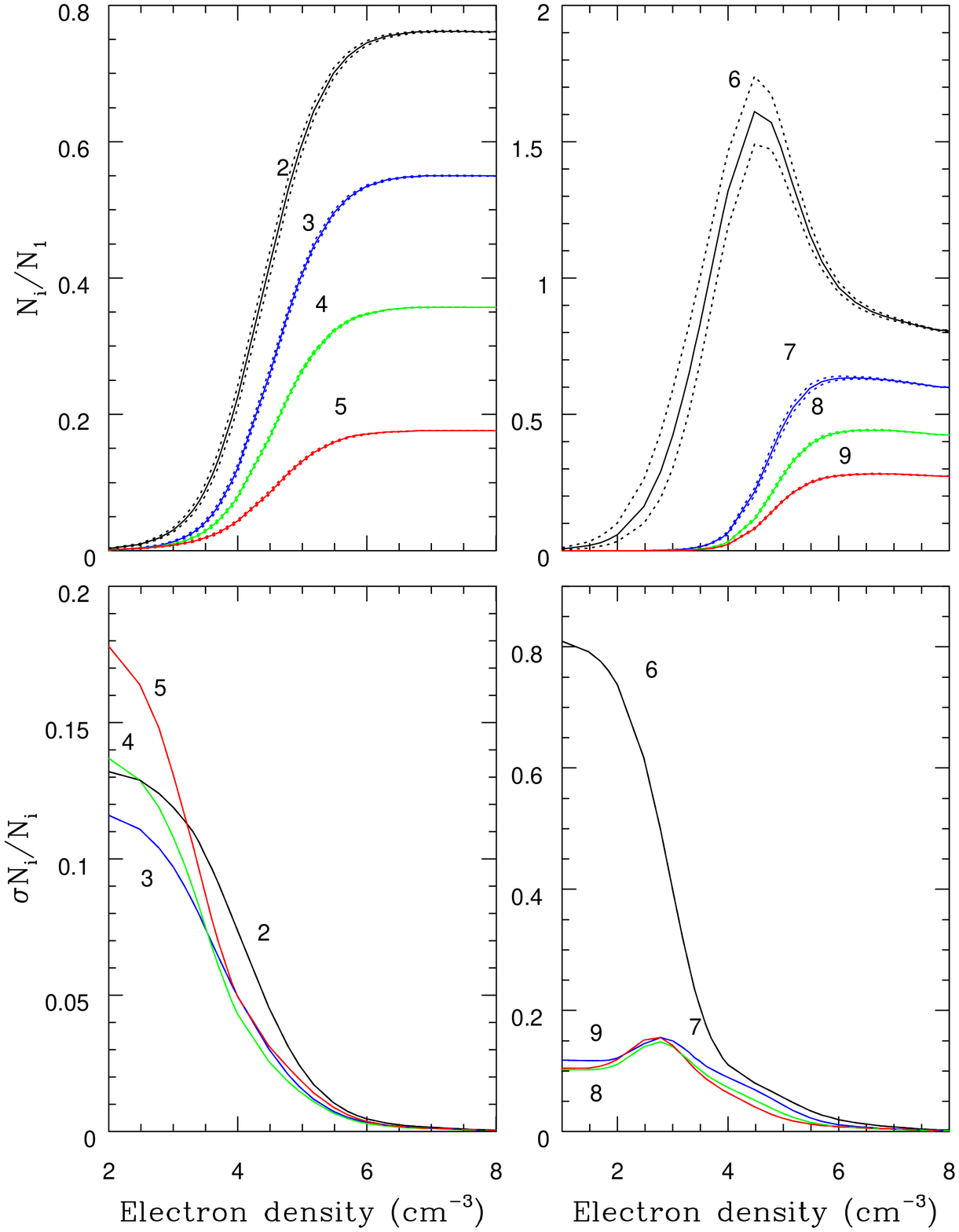}
\caption{Like \ref{errorfe2a} but from uncertainties in lifetimes and radiative yields estimated from
the dispersion among various calculations.} 
\label{errorfe2b}
\end{figure}

Figure \ref{errordist2} shows the estimated lifetime uncertainties for the lowest 52 levels of Fe II. The figure also
presents the level population uncertainties that results from the present uncertainties in lifetimes and transition
yields for a plasma with $T_e=10^4$ K and $n_e=10^4$ cm$^{-3}$. Here, the 
adopted uncertainties in the collision strengths are kept at 20\% for all 
transitions. By far, the most uncertain lifetime is that of the important 
$a\ ^4F_{9/2}$ level ($i=6$), yet
the way that this uncertainty propagates through level populations depends on the density of the plasma. For electron densities
much lower than the critical density for the level the uncertainty in the lifetime reflects directly
on the level population for that level. This is seen at $n_e=10^4$ cm$^{-3}$ for levels $\sim$18 and higher. However, as the density increases the uncertainties in the level populations become incresingly dominated
by the collision strengths. This effect is clearly illustrated in Figure \ref{errorfe2b}.

\section{Uncertainties in Emission Line Emissivities and Diagnostic Line Ratios}

The line emissivity, in units of photons per second, of a transition $i\to f$, with $i>f$, is 
\begin{equation}
j_{i,f}=N_i\times A_{i,f}.
\end{equation}\label{emisseq}
In computing the uncertainty in $j_{i,f}$ one must to account for the fact that $N_i$ and $A_{i,f}$ are
correlated, because the latter appears in the denominator term of Equation (1) that determines $N_i$. This is important
because the most frequently observed lines from any upper level are usually those that dominate the total
decay rate for the level, i.e., the inverse of the level's lifetime. 
It is convenient to re-write the above equation as
\begin{equation}
j_{i,f}=\kappa_i {A_{i,f}\over n_e\sum_j q_{i,j} + \sum_j A_{i,j}}.
\end{equation}
Combining this equation with Equation \ref{leverr} one finds 
\begin{equation}
\left({\delta j_{i,f}\over j_{i,f}}\right)^2 = 
\left({\delta N_i\over N_i}\right)^2
-\left({N_i\over \kappa_i\tau_i}\right)^2\left({\delta\tau_i \over \tau_i}\right)^2
+\left(1-{N_i\over \kappa_i}A_{i,f}\right)^2
\left({\delta A_{i,f}\over A_{i,f}}\right)^2.
\end{equation}\label{emisser} 
This equation can be readily evaluated from the level populations and uncertainties already known. The equation
has various interesting properties: (1)
the equation is independent of the physical units used for the emissivities; (2) in the high density limit, as the
uncertainty in the level population goes to zero, the uncertainty in the emissivity is the same as in the A-value.

\begin{figure*}
\epsscale{0.8}\plotone{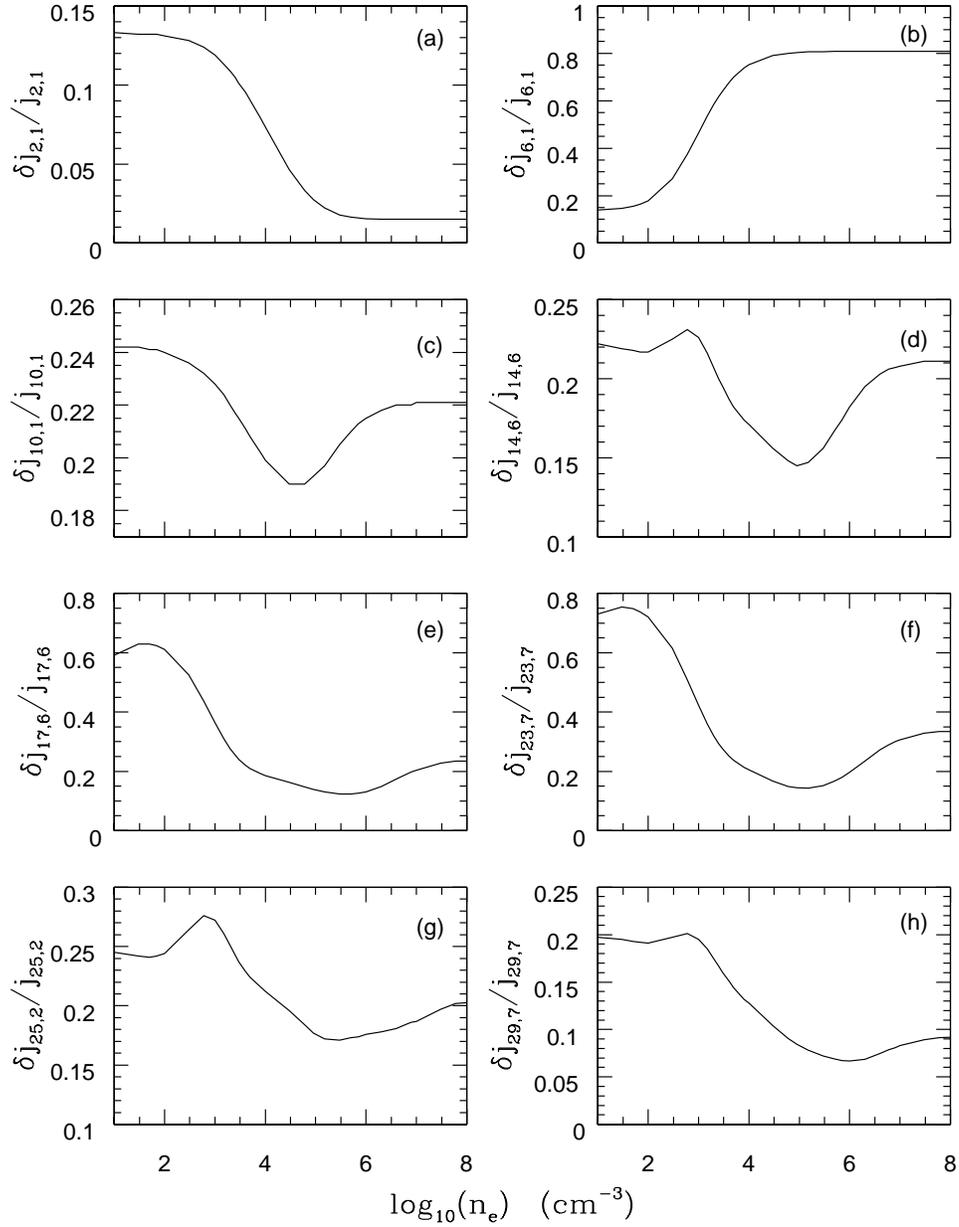}
\caption{Uncertainties in [Fe II] line emissivities at $10^4$ vs. $n_e$. The transitions shown are:
(a) 25.9 $\mu$m ($a\ ^6D_{7/2} - a\ ^6D_{9/2}$); (b) 5.33 $\mu$m ($a\ ^4F_{9/2} - a\ ^6D_{9/2}$);
(c) 1.256 $\mu$m ($a\ ^4D_{7/2} - a\ ^6D_{9/2}$); (d) 8616.8 \AA ($a\ ^4P_{5/2} - a\ ^4F_{9/2}$);
(e) 7155.2 \AA ($a\ ^2G_{9/2} - a\ ^4F_{9/2}$); (f) 5527.4 \AA ($a\ ^2D_{5/2} - a\ ^4F_{7/2}$);
(g) 4889.7 \AA ($b\ ^4P_{5/2} - a\ ^6D_{7/2}$); (h) 5261.6 \AA ($a\ ^4H_{11/2} - a\ ^4F_{7/2}$).} 
\label{lineerror}
\end{figure*}

Figure \ref{lineerror} depicts uncertainties in emissivity for a sample of strong IR, near-IR, and optical
[Fe II] lines. These are computed at $10^4$ K. The uncertainties in the 
collision strengths are 20\% and the uncertainties in the lifetimes and A-values are those estimated in the previous section.
The behavior of these uncertainties for different physical conditions is complex.
Let us look, for instance, at the uncertainty of emissivity of the 5.3$\mu$m line ($a\ ^4F_{9/2} - a\ ^6D_{9/2}$; $6\to 1$)
whose behavior is contrary to the uncertainty in the population of the $a~^4F_{9/2}$ level (see Figure \ref{errorfe2b}). 
According to equations \ref{popeq} and \ref{emisseq}, in the low density limit
\begin{equation}
j_{i,f} \to \sum_k N_k n_e q_{k,i}\left({A_{i,f}\over \sum_{j<i} A_{i,j}}\right).
\end{equation}
In the case of the $a\ ^4F_{9/2}$ level the 5.3$\mu$m transition dominates the total decay rate of level
and the ratio $A_{i,f}/\sum_{j<i} A_{i,j}$ is essentially 1. Thus, the uncertainty in the $A_{6,1}$
rate cancels out at low electron densities and the uncertainty in the emissivity is small despite a
large uncertainty in the level population. By contrast, at high densities the population of the 
level approaches the Boltzmann limit and the uncertainty in the emissivity is solely given by that in 
$A_{6,1}$, which is $\sim 80\%$.

A line emissivity ratio between two lines is given by
\begin{equation}
R = {j_{i,f}\over j_{g,h}} = \left({N_i\over N_g}\right) \left({A_{i,f}\over 
A_{g,h}}\right)\left({\Delta E_{i,f}\over \Delta E_{g,h}}\right),
\end{equation}
where $\Delta E_{i,f}$ is the energy difference between levels $i$ and $f$
and  we have used emissivties in units of energy per second.
In computing the uncertainty in this line ratio one must account for the 
fact that the emissivities are correlated.
Moreover, a general expression for the 
uncertainty must account for cases where $i=g$, in which case the uncertainty in the ratio
would depend only on the A-values.
The uncertainty is the ratio is given by
\begin{equation}
\begin{split} 
\left({\delta R \over R}\right)^2 = 
\left[1-R\left({\partial j_{g,h}\over \partial j_{i,f}}\right)\right]^2
\left({\delta j_{i,f}\over j_{i,f}}\right)^2
+\left[1-R\left({\partial j_{i,f}\over \partial j_{g,h}}\right)\right]^2
\left({\delta j_{g,h}\over j_{g,h}}\right)^2, 
\end{split}  
\end{equation}
where 
$$
{\partial \over \partial j_{i,f}} = {1\over A_{i,f}}{\partial \over
\partial N_i} + {1\over N_{i}}{\partial \over
\partial A_{i,f}}. 
$$
Thus, 
\begin{equation}
\begin{split} 
\left({\delta R \over R}\right)^2 =  \left[1-R\left({A_{g,h}\Delta E_{g,h}\over
A_{i,f}\Delta E_{i,f}}{\partial N_g\over \partial N_i} + 
{A_{g,h}\Delta_{g,h}\over \Delta E_{i,f} N_i}
{\partial N_g\over \partial A_{i,f}}\right)\right]^2\left(
{\delta j_{i,f}\over j_{i,f}}\right)^2 +
\\     
\left[1-R\left({A_{i,f}\Delta E_{i,f}\over
A_{g,h}\Delta E_{g,h}}{\partial N_i\over \partial N_g} + 
{A_{i,f}\Delta_{i,f}\over \Delta E_{g,h} N_g}
{\partial N_i\over \partial A_{g,h}}\right)\right]^2\left(
{\delta j_{g,h}\over j_{g,h}}\right)^2.
\end{split} 
\end{equation} 
From Equation \ref{popeq} we find $(\partial N_i/\partial A_{g,h})=
N_iN_g/\kappa_i$ for $h=i$, $(\partial N_i/\partial A_{g,h})=-N_i^2/\kappa_i$ for $g=i$, and 
$(\partial N_i/\partial A_{g,h})=0$ otherwise.

\begin{figure}
\epsscale{0.7}\plotone{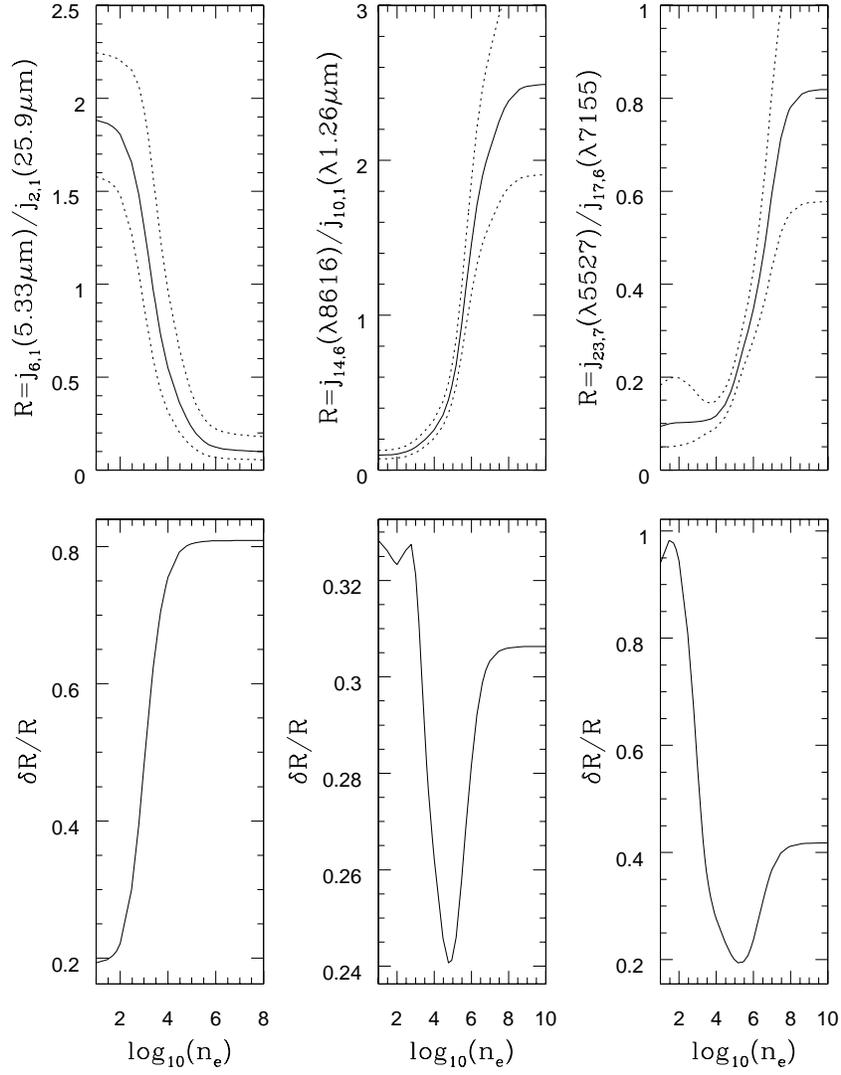}
\caption{[Fe II] emissivity line ratios (upper panel) and uncertainties (lower
panel) at $10^4$ K vs. electron density.}   
\label{raterr}
\end{figure}

In the general case of a ratio involving several lines in the numerator and/or
denominator, i.e.,
\begin{equation}
R={\sum_{\{i,f\}} j_{i,f}\over \sum_{\{g,h\}} j_{g,h} },
\end{equation}
the uncertainty is
\begin{equation}
\begin{split}  
\left({\delta R\over R}\right)^2 = \sum_{\{i,f\}} \left(
{\sum_{\{i,f\}'} (\partial j_{\{i,f\}'}/\partial j_{\{i,f\}})\over
\sum_{\{i,f\}} j_{\{i,f\}}}
- {\sum_{\{g,h\}} (\partial j_{\{g,h\}}/\partial j_{\{i,f\}})\over
\sum_{\{g,h\}} j_{\{g,h\}}}
\right)^2 (\delta j_{\{i,f\}})^2
\\     
\sum_{\{g,h\}} \left(
{\sum_{\{g,h\}'} (\partial j_{\{i,f\}'}/\partial j_{\{g,h\}})\over
\sum_{\{g,h\}} j_{\{g,h\}}}
- {\sum_{\{g,h\}'} (\partial j_{\{g,h\}'}/\partial j_{\{g,h\}})\over
\sum_{\{g,h\}} j_{\{g,h\}}} \right)^2 (\delta j_{\{g,h\}})^2
\end{split}  
\end{equation}

Figure \ref{raterr} shows a sample of line ratios between IR and optical lines
and their uncertainties. The uncertainties exhibit complex behaviour with
changes in density and temperatures. 
In general, line ratios are only useful as diagnostics when the observed
ratio lies around middle range of the theoretical ratio. 
Moreover, it is very important to know the uncertainties in the ratios
when selecting appropriate diagnostics from a given spectrum.

\section{Conclusions}

We presented a method to compute uncertainties in spectral models from
uncertainties in atomic/molecular data. 
Our method is very efficient and allows us to compute uncertainties in
all level populations by solving a single algebraic equation.
Specifically, we treat the case
of non-LTE models where electron impact excitation is balanced by
spontaneous radiative decay. However, the method can be extended to 
ionization balance and additional excitation mechanisms. 

Our method is tested in O III and Fe II models, first by assuming
commonly assumed uncertainties and then by adopting uncertainties in 
lifetimes and A-values given by the dispersion between the results of
multiple independent computations. Moreover, we show that uncertainties
taken this way are in practice very good estimates.

Then we derive analytic expresions for the uncertainties in line emissivities and line ratios. 
These equations take into account the correlations between 
level populations and line emissivities.
Interestingly, the behaviour of uncertainties in level populations and 
uncertainties in emissivities for transitions from the same upper levels
are often different and even opposite. This is the case, in particular, 
for lines that result from transitions that dominate the total dacay rate
of the upper level. Then, the uncertainties in A-values for the transitions
that yield the lines cancel out with the uncertainties in the 
lifetimes of the levels.
In terms of emission line ratios, 
it is also found that knowledge of the uncertainties in the ratios
is essential selecting appropriate ratios for density and temperature
diagnostics. 

At present, we are in the process of estimating uncertainties in atomic
data for species of astronomical interest. Our uncertainty estimates and analysis of
the uncertainties in various spectral models, ionic abundance
determinations, and dianostic line ratios will be presented in 
future publicaitons.
%
%==================================================================================
\acknowledgments
%
%
%==================================================================================
%
%\bibliographystyle{apj}
%\bibliography{my-references}

\begin{acknowledgements}
We acknowledge financial support from grants
from the NASA Astronomy and Physics Research and Analysis Program (award NNX09AB99G).
The Belgian FRS-FNRS is also acknowledged. 
\end{acknowledgements}

%%%% References

%
%==================================================================================
%
%
%==================================================================================
%
%
\end{document}